\documentclass{INTERSPEECH2023}

\interspeechcameraready

\usepackage{multicol}
\usepackage{multirow}
\usepackage{color, colortbl}
\usepackage{xcolor}
\colorlet{eng}{blue!10}
\colorlet{cmn}{teal!10}
\colorlet{multi}{yellow!10}
\colorlet{euro}{orange!10}

\title{ML-SUPERB: Multilingual Speech Universal PERformance Benchmark}
\name{Jiatong Shi$^1$, Dan Berrebbi$^1$\sthanks{Equal contribution, sorted in alphabetical order.}, William Chen$^{1*}$, Ho-Lam Chung$^{2*}$, En-Pei Hu$^{2*}$, Wei Ping Huang$^{2*}$, Xuankai Chang$^1$, Shang-Wen Li$^3$, Abdelrahman Mohamed$^4$, Hung-yi Lee$^2$, Shinji Watanabe$^1$}
\address{
  $^1$Carnegie Mellon University \quad $^2$National Taiwan University \quad $^3$Meta AI \quad $^4$Rembrand}
\email{\{jiatongs, dberrebbi, wc4, swatanab\}@cs.cmu.edu, shangwenl@meta.com, abdo@rembrand.com hungyilee@ntu.edu.tw}

\usepackage[
backend=biber,
style=ieee,
citestyle=numeric-comp,
maxbibnames=3,
maxcitenames=3,
doi=false,isbn=false,url=false,eprint=false
]{biblatex}

\defbibheading{bibliography}[\refname]{}

\DeclareSourcemap{
	\maps[datatype=bibtex, overwrite=true]{
		\map{
			\step[fieldsource=booktitle,
			match=\regexp{.*Interspeech.*},
			replace={Proc. Interspeech}]
			\step[fieldsource=journal,
			match=\regexp{.*INTERSPEECH.*},
			replace={Proc. Interspeech}]
			\step[fieldsource=booktitle,
			match=\regexp{.*ICASSP.*},
			replace={Proc. ICASSP}]
			\step[fieldsource=booktitle,
			match=\regexp{.*icassp_inpress.*},
			replace={Proc. ICASSP (in press)}]
			\step[fieldsource=booktitle,
			match=\regexp{.*Acoustics,.*Speech.*and.*Signal.*Processing.*},
			replace={Proc. ICASSP}]
			\step[fieldsource=booktitle,
			match=\regexp{.*International.*Conference.*on.*Learning.*Representations.*},
			replace={Proc. ICLR}]
			\step[fieldsource=booktitle,
			match=\regexp{.*International.*Conference.*on.*Computational.*Linguistics.*},
			replace={Proc. COLING}]
			\step[fieldsource=booktitle,
			match=\regexp{.*SIGdial.*Meeting.*on.*Discourse.*and.*Dialogue.*},
			replace={Proc. SIGDIAL}]
			\step[fieldsource=booktitle,
			match=\regexp{.*International.*Conference.*on.*Machine.*Learning.*},
			replace={Proc. ICML}]
			\step[fieldsource=booktitle,
			match=\regexp{.*North.*American.*Chapter.*of.*the.*Association.*for.*Computational.*Linguistics:.*Human.*Language.*Technologies.*},
			replace={Proc. NAACL}]
			\step[fieldsource=booktitle,
			match=\regexp{.*Empirical.*Methods.*in.*Natural.*Language.*Processing.*},
			replace={Proc. EMNLP}]
			\step[fieldsource=booktitle,
			match=\regexp{.*Association.*for.*Computational.*Linguistics.*},
			replace={Proc. ACL}]
			\step[fieldsource=booktitle,
			match=\regexp{.*Automatic.*Speech.*Recognition.*and.*Understanding.*},
			replace={Proc. ASRU}]
			\step[fieldsource=booktitle,
			match=\regexp{.*Spoken.*Language.*Technology.*},
			replace={Proc. SLT}]
			\step[fieldsource=booktitle,
			match=\regexp{.*Speech.*Synthesis.*Workshop.*},
			replace={Proc. SSW}]
			\step[fieldsource=booktitle,
			match=\regexp{.*workshop.*on.*speech.*synthesis.*},
			replace={Proc. SSW}]
			\step[fieldsource=booktitle,
			match=\regexp{.*Advances.*in.*neural.*information.*processing.*},
			replace={Proc. NeurIPS}]
			\step[fieldsource=booktitle,
			match=\regexp{.*Advances.*in.*Neural.*Information.*Processing.*},
			replace={Proc. NeurIPS}]
			\step[fieldsource=booktitle,
			match=\regexp{.*Workshop.*on.* Applications.* of.* Signal.*Processing.*to.*Audio.*and.*Acoustics.*},
			replace={Proc. WASPAA}]
			\step[fieldsource=publisher,
			match=\regexp{.+},
			replace={{}}]
			\step[fieldsource=month,
			match=\regexp{.+},
			replace={{}}]
			\step[fieldsource=location,
			match=\regexp{.+},
			replace={{}}]
			\step[fieldsource=address,
			match=\regexp{.+},
			replace={{}}]
			\step[fieldsource=organization,
			match=\regexp{.+},
			replace={{}}]
		}
	}
}

\begin{document}

\maketitle
 
\begin{abstract}
Speech processing Universal PERformance Benchmark (SUPERB) is a leaderboard to benchmark the performance of Self-Supervised Learning (SSL) models on various speech processing tasks. However, SUPERB largely considers English speech in its evaluation. This paper presents multilingual SUPERB (ML-SUPERB), covering 143 languages (ranging from high-resource to endangered), and considering both automatic speech recognition and language identification. Following the concept of SUPERB, ML-SUPERB utilizes frozen SSL features and employs a simple framework for multilingual tasks by learning a shallow downstream model. Similar to the SUPERB benchmark, we find speech SSL models can significantly improve performance compared to FBANK features. Furthermore, we find that multilingual models do not always perform better than their monolingual counterparts. We will release ML-SUPERB as a challenge with organized datasets and reproducible training scripts for future multilingual representation research.
\end{abstract}
\noindent\textbf{Index Terms}: speech self-supervised learning, multilingual speech recognition, language identification

\section{Introduction}

% \begin{itemize}
%     \item the brief summary of the multilingual superb benchmark
%     \item Differences from XTREME-S  (Self-supervised v.s. unconstrained; coverage of languages; perspective of concentration/evaluation; small scale (low-resources/few-shot learning); system design)
% \end{itemize}

Self-supervised learning (SSL) has been a popular method in the speech community. 
SSL models have shown promising results by capturing important speech features, such as phonemes and other acoustic units, through training on large amounts of unlabeled speech data \cite{mohamed2022self}. These models have led to significant improvements in downstream tasks, such as speech recognition, speaker identification, and emotion recognition \cite{yang21c_interspeech}. Over the past few years, researchers have proposed a variety of SSL models with different training objectives, operating under various data conditions, model architectures, and modalities \cite{baevski2020wav2vec, hsu2021hubert}.

% A common issue in speech SSL models is the difficulty of model comparison, as most of the SSL models were evaluated with different experimental setups. To address the issue, Yang et al. release Speech processing Universal PERformance Benchmark (SUPERB) \cite{yang21c_interspeech}. With its extension SUPERB-SG \cite{tsai2022superb}, SUPERB now supports a comprehensive speech SSL benchmark with tasks of recognition, detection, semantics, speaker, paralinguistics, and generation.

A major challenge in evaluating SSL models for speech is the difficulty of comparison since most models have been evaluated using different experimental setups. To address this issue, Yang et al. introduced the Speech processing Universal PERformance Benchmark (SUPERB) \cite{yang21c_interspeech}. Recently, an extension of SUPERB called SUPERB-SG \cite{tsai2022superb} has been introduced. SUPERB provides a comprehensive speech SSL benchmark including tasks such as recognition, detection, semantics, speaker identification, paralinguistics, and generation. With SUPERB, researchers can more easily compare the performance of different SSL models on various speech-related tasks, universally.

% Though covering various speech tasks, SUPERB in its design only concentrates on \textit{only} English speech. However, we observe an increasing interest in applying SSL models to multilingual scenarios, such as learning multilingual SSL models \cite{babu2021xls, conneau2020unsupervised, duquenne2022speechmatrix} or applying SSL models in a cross-lingual manner \cite{zhao2022improving, berrebbi22_interspeech, wu2020self, alghezi21_interspeech}. Considering those application scenarios, we propose multilingual SUPERB (ML-SUPERB) in this paper to benchmark and support the analysis of SSL in future multilingual studies.

While SUPERB covers a wide range of speech tasks, it was designed primarily for English speech. However, there has been growing interest in applying SSL models to multilingual scenarios, such as training multilingual SSL models \cite{babu2021xls, conneau2020unsupervised, duquenne2022speechmatrix} or using SSL models in a cross-lingual manner \cite{zhao2022improving, berrebbi22_interspeech, wu2020self,  li22aa_interspeech}. To support future research in these areas, we propose a new benchmark called multilingual SUPERB (ML-SUPERB). % This benchmark will provide a framework for evaluating and analyzing SSL models in multilingual settings.

% ML-SUPERB covers 143 languages, ranging from high-resource languages (e.g., English) to endangered languages (e.g., Totonac). For evaluation tasks, ML-SUPERB mainly focuses on  automatic speech recognition (ASR) and language identification (LID). Following common applications of SSL models, the benchmark can be further split into four tracks: \textit{single language ASR track}, \textit{multilingual ASR track}, \textit{LID track}, \textit{joint multilingual ASR/LID track}. Similar to SUPERB, for training efficiency, ML-SUPERB uses frozen SSL models as representation extractors and a lightweight downstream model fine-tuned for different tracks.

ML-SUPERB is designed to cover a wide range of languages, including both high-resource languages like English and endangered languages such as Totonac. The benchmark primarily focuses on evaluating SSL models for automatic speech recognition (ASR) and language identification (LID). To accommodate different use cases for SSL models, ML-SUPERB includes two tracks with four different tasks: the monolingual track (monolingual ASR), and the multilingual track (multilingual ASR, LID, joint multilingual ASR/LID). Similar to SUPERB, ML-SUPERB employs frozen SSL models as feature extractors and a lightweight downstream model that can be fine-tuned for different tracks to achieve high training efficiency.

% There are some existing benchmarks that also consider multilingual SSL benchmarks \cite{evain21_interspeech, conneau22_interspeech}. Lebenchmark considers multilingual speech SSL models in their benchmark, but only holds speech tasks in French \cite{evain21_interspeech}.  XTREME-S focuses on multilingual speech representation benchmarks with tasks of ASR, speech translation, speech classification, and speech retrieval\cite{conneau22_interspeech}. Three major aspects are different between XTREME-S and ML-SUPERB: (1) language coverage: 102 languages are studied in XTREME-S, while 143 are benchmarked in ML-SUPERB; (2) task coverage: 4 tasks are considered in XTREME-S, while ML-SUPERB focuses only on ASR and LID. In ASR, ML-SUPERB involves four tracks to evaluate the common scenarios in multilingual research, while XTREME-S focuses only on multilingual ASR. (3) scale difference: ML-SUPERB generally focuses on a smaller scale of experiments for efficiency purposes, including the use of small benchmark datasets and lightweight downstream models. Meanwhile, we do not consider fine-tuning in ML-SUPERB.

Several existing benchmarks also include multilingual SSL models \cite{evain21_interspeech, javed2022indicsuperb, conneau22_interspeech}. Lebenchmark primarily evaluates speech tasks in French \cite{evain21_interspeech}; IndicSUPERB focuses mostly on Indian languages \cite{javed2022indicsuperb}. XTREME-S focuses on multilingual speech representation benchmarks, including ASR, speech translation, speech classification, and speech retrieval \cite{conneau22_interspeech}. There are three main differences between XTREME-S and ML-SUPERB. Firstly, ML-SUPERB covers a wider range of languages, with 143 languages compared to XTREME-S's 102. Secondly, ML-SUPERB focuses on ASR and LID, while XTREME-S covers four different tasks. However, ML-SUPERB expands the tasks by evaluating them in four common multilingual research scenarios, while XTREME-S considers multilingual training only. Finally, ML-SUPERB is designed for efficiency, using smaller benchmark datasets and downstream models, and does not include fine-tuning. This lightweight setup allows us to conduct experiments for a dozen of popular speech SSL models, trained with various sizes and pre-training sets, and compare their performances across the proposed tracks. We expect ML-SUPERB would be a valuable complement to existing benchmarks.

\section{Benchmark Details}

% \begin{figure}[tbp]
%     \centering
%     \includegraphics[width=\linewidth]{mSUPERB_framework.pdf}
%     \caption{ML-SUPERB design. ML-SUPERB corpus is discussed in Sec.~\ref{ssec: data} and ML-SUPERB tracks are introduced in Sec.~\ref{ssec: single lang track} and Sec.~\ref{ssec: masr track}.}
%     \label{fig:diagram}
%         \vspace{-15pt}
% \end{figure}

\begin{table*}[t]
    \centering
    \caption{Statistics of the data used for training, development, and testing in ML-SUPERB. Detailed discussed in Sec.~\ref{ssec: data}.}
    \vspace{-10pt}
    % \resizebox {\linewidth} {!} {

%     \begin{tabular}{l|c|cc}
%         \toprule
% Dataset & Size(h) & Normal & Few-shot \\
% \midrule
% 10-minute & 37.43 & $\sim$10min $\times$ 123 langs & 5 utt. $\times$ 20 langs \\
% 1-hour & 222.46 & $\sim$1h $\times$ 123 langs & 5 utt. $\times$ 20 langs \\
% Development  & 41.82 & $\sim$10min $\times$ 123 langs & $\sim$10min $\times$ 20 langs \\
% Test  & 44.97 & $\sim$10min $\times$ 123 langs & $\sim$10min $\times$ 20 langs \\
%         \bottomrule
%     \end{tabular}

% }

% \resizebox {\linewidth} {!} {

    \begin{tabular}{l|c|cc}
        \toprule
Dataset & Hours & Normal Langs (123) & Few-shot Langs (20)\\
\midrule
10-minute & 37.43 & $\sim$10min $\times$ 240 (\texttt{lang}, \texttt{data}) & 5 utt. $\times$ 20 \texttt{lang} \\
1-hour & 222.46 & $\sim$1h $\times$ 240 (\texttt{lang}, \texttt{data}) & 5 utt. $\times$ 20 \texttt{lang}\\
Dev  & 41.82 & $\sim$10min $\times$ 240 (\texttt{lang}, \texttt{data}) & $\sim$10min $\times$ 31 (\texttt{lang}, \texttt{data}) \\
Test  & 44.97 & $\sim$10min $\times$ 240 (\texttt{lang}, \texttt{data}) & $\sim$10min $\times$ 31 (\texttt{lang}, \texttt{data}) \\
        \bottomrule
    \end{tabular}

% }
    \label{tab: dataset}
    \vspace{-20pt}
\end{table*}

\subsection{Data Collection}
\label{ssec: data}

ML-SUPERB gathers data from a wide range of multilingual speech corpora, including Multilingual Librispeech \cite{pratap2020mls}, Commonvoice \cite{ardila2020common}, Voxforge \cite{maclean2018voxforge}, Voxpopuli \cite{wang2021voxpopuli}, Googlei18n open-source project \cite{kjartansson-etal-tts-sltu2018, kjartansson-etal-2020-open,he-etal-2020-open}, Nordic Language Technology ASR corpora \cite{rehm2012language}, Fleurs \cite{conneau2023fleurs}, NCHLT Speech \cite{barnard2014nchlt}, Spoken Wikipedia corpus \cite{baumann2019spoken}, Mexican endangered languages \cite{shi2021leveraging,shi2021highland,berrebbi22_interspeech, amith_yoloxochitl_mixtec, amith_audio_corpus_sierra, amith_totonac}, M-AILab multilingual corpora \cite{mailab}, Living Audio dataset \cite{braude2019all}, ALFFA corpus \cite{de2014smartphone}. All corpora are with either Creative Commons, MIT, GNU, or Free-BSD licenses, which are available for both industrial and academic research, permissively.

For each language-corpus pair denoted as (\texttt{lang}, \texttt{data}), three 10-minute subsets are randomly extracted for training, development, and testing, along with an additional 1-hour training set that includes the 10-minute training set.\footnote{We used the original split for source datasets, with the exception of SWC, M-AILABS, LAD, and ALFFA. Therefore, all datasets except these four can be used for SSL pre-training.} The reasons for using a small 10-minute/1-hour training set are: (1) \textit{Challenging design}: using a large training data size could lead to high performance easily and may result in a saturated benchmark in evaluation metrics \cite{baevski2020wav2vec, hsu2021hubert}. Therefore, using a smaller training set size presents a more challenging design for the SSL models, which can help evaluate their robustness and generalization capability.
 (2) \textit{Reasonable performance}: previous speech SSL works have frequently adopted 10-minute and 1-hour training sizes. Even in such extreme cases, the performances with SSL are generally reasonable  \cite{baevski2020wav2vec, hsu2021hubert}, indicating that this setting could be a feasible solution to the benchmark as well.
 (3) \textit{Training efficiency}: with 143 languages coverage, limiting the training size is important to keep the experiments within reasonable computational efforts. Using a smaller training set size can help reduce the computational cost and make the training process more efficient. A full evaluation cycle of ML-SUPERB can take up to 3 days using 4 2080Ti GPUs.

% We choose to use a small 10-minute/1-hour training set for three reasons: (1) challenging design. It has been found that with a large training data size, SSL models could reach high performance easily \cite{baevski2020wav2vec, hsu2021hubert}, which may lead to a saturated benchmark in evaluation metrics. (2) reasonable performance. 10-minute and 1-hour training sizes have been frequently adopted in previous speech SSL works \cite{baevski2020wav2vec, hsu2021hubert}. Even in such extreme cases, the performances with SSL are generally reasonable, so we believe the setting could be a feasible solution to the benchmark as well. (3) training efficiency. Considering the 143 languages coverage, limiting the training set size is important to keep the experiments running with reasonable computational efforts.

Additionally, the benchmark includes few-shot cases with 20 languages and uses only 5 utterances in training for each language. These reserved few-shot training sets are not used in the monolingual ASR track. A detailed summary of the dataset is shown in Table~\ref{tab: dataset}.

\subsection{Monolingual Track}
\label{ssec: single lang track}

The literature suggests that speech SSL models are commonly fine-tuned on monolingual corpora \cite{zhao2022improving, berrebbi22_interspeech, wu2020self}. In ML-SUPERB, we introduce a dedicated track for monolingual ASR to facilitate this approach. We select nine languages based on geographical and linguistic considerations to balance language and domain coverage with manageable experimental mass.% \footnote{The nine languages include \texttt{eng}, \texttt{deu}, \texttt{fra}, \texttt{rus}, \texttt{swa}, \texttt{swe}, \texttt{jpn}, \texttt{cmn}, and \texttt{xty} in ISO-639-3 codes}

In total, we introduce 14 \texttt{monolingual\_exp}.
For a \texttt{monolingual\_exp} in language \texttt{lang} we select one dataset of this language and use it for training the model and for validation\footnote{Each \texttt{monolingual\_exp} is made of one experiment with the 10-minute set for training and one with the 1-hour set.}. 
For evaluation of a \texttt{monolingual\_exp}, we use all the datasets of \texttt{lang} to test the trained model on various accent or domain conditions.
We select one pair (\texttt{lang} , \texttt{data}) for training for \texttt{lang}$\in$ \{\texttt{rus}, \texttt{swa}, \texttt{swe}, \texttt{jpn}, \texttt{cmn}, \texttt{xty}\}.
For \texttt{lang}$\in$ \{\texttt{eng}, \texttt{fra}, \texttt{deu}\} we select respectively 3, 2 and 2 pairs (\texttt{lang}, \texttt{data}) in order to evaluate the impact of the training domain on the models' performances. For instance, for \texttt{eng} we have 3 \texttt{monolingual\_exp}, with (\texttt{eng} , \texttt{MLS}), (\texttt{eng} , \texttt{NCHLT}) and (\texttt{eng} , \texttt{VoxPopuli}).

\subsection{Multilingual Track}
\label{ssec: masr track}

\noindent \textbf{Multilingual ASR task}: in the multilingual ASR task, we use the training set where combining text transcriptions from all 143 languages.
The multilingual ASR task has two sub-tasks on the 10-minute train set and the 1-hour train set. For both training sets, we reserve 20 languages for few-shot learning scenarios as discussed in Sec.~\ref{ssec: data}. In this track, the model is expected to directly predict the correct orthography in the target language.

\noindent \textbf{LID task}: LID track focuses on language identification with the same training set of 143 languages in 10 minutes and 1 hour. However, we do not consider evaluation for languages with few-shot settings, given that the identification of those languages is very challenging due to the label biasing.

\noindent \textbf{Joint Multilingual ASR/LID task}: A widely used technique in previous literature involves adding the language ID to the start of the speech transcript to facilitate joint training of multilingual ASR and LID models \cite{watanabe2017language, hou20_interspeech, zhang22da_interspeech, chen2023improving}. Joint training can improve performance in certain scenarios, and it can also enhance model interpretability by separating language identification errors. Therefore, we have included this task in our multilingual track. The task's design is the same as the multilingual ASR task for ASR and the LID task for language identification.

\subsection{Framework and Benchmark Settings}
\label{ssec: framework}

\noindent \textbf{Toolkits}: We utilize the S3PRL toolkit \cite{yang21c_interspeech} for upstream models, which offers a wide range of speech SSL model architectures and APIs that support customized SSL models from Huggingface \cite{wolf-etal-2020-transformers} and user-defined models. For task-specific downstream training, we use ESPnet \cite{watanabe2018espnet}. We plan to publish ML-SUPERB as an all-in-one recipe in ESPnet's \texttt{egs2} recipe collection, encompassing data preprocessing, training, inference, and evaluation\footnote{\url{https://github.com/espnet/espnet/tree/master/egs2/ml_superb/asr1}}.

\noindent \textbf{Downstream model and training details}: Our downstream model design is based on the SUPERB concept. First, we compute a weighted summation of frozen speech SSL representations using learnable weights. Next, we apply a convolutional downsample layer that reduces the sequence of speech SSL features by half, passing the resulting hidden states to a transformer model consisting of two layers with an attention dimension of 256, a feedforward layer dimension of 1024, and 8 attention heads. A dropout rate of 0.1 is employed, and the model is trained using the connectionist temporal Ccassification loss. We use the Adam optimizer with a learning rate of 0.0001 and 1e-6 weight decay. Specaugment is applied to the representation (i.e., the weighted sum of speech SSL representation) following the SUPERB benchmark. The batch size is set to 8 with the gradient accumulation as 4. The same configuration is used for all tasks in both the monolingual and multilingual tracks.

The number of iterations in training is the only difference across tasks. In the monolingual track, due to the small training size, we set it to 15,000. In the multilingual track, we use 300,000 iterations for the 10-minute train set and 600,000 for the 1-hour train set. 

\noindent \textbf{Evaluation metric}: In the monolingual track, the phoneme error rate is used for \texttt{jpn} and \texttt{cmn}, while Character Error Rate (CER) is used for the remaining languages. In the multilingual track, we use CER for ASR evaluation and accuracy rate for LID evaluation, reporting results separately for the normal training set and the few-shot training set. 

For overall performance, we use the SUPERB$_s$ metric from the SUPERB benchmark \cite{feng2023superb}. We denote $s_{t,i}(u)$ as the $i^{\text{th}}$ metrics for task~$t$ and SSL model~$u$. $T$ is the set of four tasks and $I_t$ is the set of metrics for the task $t$. SUPERB$_s$ aggregates all task-specific scores $s_t(u)$ with respect to baseline (i.e., FBANK) and state-of-the-art (SOTA) model\footnote{The SOTA models for each setting are discussed in Sec.~\ref{ssec: exp result}.} on the task~$t$. The SUPERB$_s$ is defined as:
\begin{equation}
    \resizebox{\hsize}{!}{%
    $\text{SUPERB}_s(u) = \frac{1000}{|T|}\sum_t^T\frac{1}{|I_t|}\sum_{i}^{I_t} \frac{s_{t,i}(u) - s_{t,i}(\text{FBANK})}{s_{t, i}(\text{SOTA}) - s_{t,i}(\text{FBANK})}$
    }
\end{equation}
We expect SUPERB$_s$ can provide a comprehensive view of the model performance on the benchmark and take the difficulty of tasks into consideration.

\noindent \textbf{Analysis support}: To facilitate a more comprehensive analysis of the benchmark, we provide various analysis tools. For the multilingual ASR evaluation, we present the character error rate (CER) for each language as well as aggregated scores for different language groups, in addition to the average CER for both normal and few-shot cases. In line with previous studies \cite{chang2021exploration, chen2022wavlm}, we also offer visualizations of the learnable layer weights and their learning curve during training.

\section{Experiments}

\subsection{Candidate models}
\label{ssec: models}

\begin{table}[t]
    \centering
    \caption{Description of the candidate models.}
    \vspace{-10pt}
    \resizebox {\linewidth} {!} {

    \begin{tabular}{l|c|cc}
        \toprule
        \multirow{2}{*}{Model} & \multirow{2}{*}{Params (M)} &  \multicolumn{2}{c}{Pre-Training}   \\
         &  &  \# Hours &  \# Langs \\
\midrule
\rowcolor{eng} wav2vec2-base \cite{baevski2020wav2vec}  & 95 & 1k & 1 \\
\rowcolor{eng} wav2vec2-large \cite{baevski2020wav2vec}  & 317 & 60k & 1 \\
\rowcolor{eng} robust-wav2vec2-large \cite{hsu21_interspeech}  & 317 & 65k & 1\\
\rowcolor{euro} wav2vec2-base-23 \cite{wang2021voxpopuli}  & 95 & 100k & 23 \\
\rowcolor{euro} wav2vec2-large-23 \cite{wang2021voxpopuli} & 317 & 100k & 23 \\
\rowcolor{multi} XLSR-53 \cite{conneau2020unsupervised}  & 317 & 56k & 53 \\
\rowcolor{multi} XLSR-128 \cite{babu2021xls}  & 317  & 400k & 128 \\
        \midrule
\rowcolor{eng} HuBERT-base \cite{hsu2021hubert} & 95 & 1k & 1  \\
\rowcolor{eng} HuBERT-large \cite{hsu2021hubert}  & 317 & 60k & 1 \\
\rowcolor{cmn} HuBERT-base-cmn \cite{mandarin_hubert}  & 95 & 10k & 1\\
\rowcolor{cmn} HuBERT-large-cmn \cite{mandarin_hubert}  & 317 & 10k & 1 \\
\rowcolor{euro} mHuBERT-base \cite{lee2022textless}  & 95 & 14k & 3 \\

        \bottomrule
    \end{tabular}

}
    \label{tab: model}
    \vspace{-15pt}
\end{table}

ML-SUPERB welcomes all speech SSL models trained on either monolingual or multilingual data. We believe the analysis of multilingual scenarios for monolingual speech SSLs is also valuable according to previous works \cite{zhao2022improving, berrebbi22_interspeech, wu2020self}. In this paper, we show the experimental results of some example model candidates as shown in Table~\ref{tab: model}.

\noindent \textbf{wav2vec2}: wav2vec2 is a popular speech SSL model for speech recognition \cite{baevski2020wav2vec}. Its pre-training uses a contrastive learning approach that prioritizes identifying true quantized latent speech representations over masked time steps from distractors. The wav2vec2 model has also been extended to many other versions for specialized use cases. For example, robust-wav2vec2-large \cite{hsu21_interspeech} considers the diversity of speech types, such as read speech, conversational speech, and noisy speech, by including additional corpora in the pre-training stage. Wav2vec2-base-23 and wav2vec2-large-23 are pre-trained on Voxpopuli \cite{wang2021voxpopuli}, with a focus on European languages. Additionally, XLSR scales up the multilingual training in wav2vec2 by incorporating more languages and data \cite{conneau2020unsupervised, babu2021xls}.

\noindent \textbf{HuBERT}: HuBERT uses an iterative offline clustering step to generate pseudo labels for each frame. During training, it predicts the pseudo labels of the masked frame, which helps to improve the quality of the learned features. Similar to wav2vec2, HuBERT also has different versions, such as a multilingual HuBERT~\cite{lee2022textless} trained in three European languages (\texttt{fra}, \texttt{spa}, \texttt{eng}) and HuBERT trained on Mandarin~\cite{mandarin_hubert}.

\begin{table*}[t]
    \centering
    \caption{10-minute set ML-SUPERB benchmark.}
            \vspace{-10pt}
    \resizebox {0.95\linewidth} {!} {
\begin{tabular}{l|c|cc|c|ccc|c}
\toprule
\multirow{3}{*}{SSL} & Monolingual ASR & \multicolumn{2}{c|}{Multilingual ASR} & \multicolumn{1}{c|}{LID} & \multicolumn{3}{c|}{Multilingual ASR + LID} & \multirow{3}{*}{SUPERB$_{s}$} \\
&          &             Normal & Few-shot & Normal & \multicolumn{2}{c}{Normal} & \multicolumn{1}{c|}{Few-shot} \\
& CER/PER & CER & CER & ACC & ACC & CER & \multicolumn{1}{c|}{CER} &  \\
\midrule
FBANK & 72.1 & 62.4 & 58.3 & 11.11 & 35.9 & 62.0 & 58.9 & 0 \\
\rowcolor{eng} wav2vec2-base \cite{baevski2020wav2vec} & 44.2 & 43.0 & 45.7 & 54.4 & 66.9 & 40.6 & 44.2 & 755.2 \\
\rowcolor{eng} wav2vec2-large \cite{baevski2020wav2vec} & 42.0 & 42.6 & 45.8 &  30.9 & 54.6 &45.5 & 50.3 & 598.3  \\
\rowcolor{eng} robust-wav2vec2-large \cite{hsu21_interspeech} & 44.4 & 40.1 & 45.4 & 50.8 & 33.1 & 38.6 & 44.9 & 680.3  \\
\rowcolor{euro} wav2vec2-base-23 \cite{wang2021voxpopuli} & 49.2 & 37.7 & 43.4 & 58.7 & 45.1 & 37.2 & 44.3 & 735.7 \\
\rowcolor{euro} wav2vec2-large-23 \cite{wang2021voxpopuli} & 42.0 & 42.1 & 44.3 & 1.1 & 21.8 & 43.4 & 46.1 & 433.8 \\
\rowcolor{multi} XLSR-53 \cite{conneau2020unsupervised} & 49.5 & 33.9 & 43.6 & 6.6 & 45.6 & 33.4 & 43.2 & 528.8 \\
\rowcolor{multi} XLSR-128 \cite{babu2021xls} & 39.7 & \textbf{29.2} & \textbf{40.9} & \textbf{66.9} & 55.6 & \textbf{28.4} & \textbf{42.1} & \textbf{947.5} \\
\rowcolor{eng} HuBERT-base \cite{hsu2021hubert} & 42.8 & 39.8 & 44.5 & 61.2 & 71.5 &  39.2 & 43.8 & 831.9 \\
\rowcolor{eng} HuBERT-large \cite{hsu2021hubert} & \textbf{38.2} & 44.4 & 48.2 &  46.5 & 55.4 & 45.6 & 49.3 & 678.7  \\
\rowcolor{cmn} HuBERT-base-cmn \cite{mandarin_hubert} & 43.1 & 40.8 & 45.4 & 49.3 & \textbf{75.1} & 37.7 & 43.5 & 779.0  \\
\rowcolor{cmn} HuBERT-large-cmn \cite{mandarin_hubert} & 39.4 & 42.6 & 45.8 & 39.5 & 66.4 & 41.9 & 45.2 & 715.4 \\
\rowcolor{euro} mHuBERT-base \cite{lee2022textless} & 41.0 & 40.5 & 45.6 & 52.4  & 46.6 & 36.8 & 44.2 &  746.2\\
\bottomrule
\end{tabular}
}

    \label{tab: 10min}
        % \vspace{-15pt}
\end{table*}

\begin{table*}[t]
    \centering
    \caption{1-hour set ML-SUPERB benchmark.}
            \vspace{-10pt}
    \resizebox {0.95\linewidth} {!} {
\begin{tabular}{l|c|cc|c|ccc|c}
\toprule
\multirow{3}{*}{SSL} & Monolingual ASR & \multicolumn{2}{c|}{Multilingual ASR} & \multicolumn{1}{c|}{LID} & \multicolumn{3}{c|}{Multilingual ASR + LID} & \multirow{3}{*}{SUPERB$_{s}$} \\
&          &             Normal & Few-shot & Normal & \multicolumn{2}{c}{Normal} & \multicolumn{1}{c|}{Few-shot} \\
& CER/PER & CER & CER & ACC & ACC & CER & \multicolumn{1}{c|}{CER} &  \\
\midrule
FBANK & 63.7 & 59.3 & 57.4 & 9.3  & 43.5 & 58.6 & 58.1 & 0 \\
\rowcolor{eng} wav2vec2-base \cite{baevski2020wav2vec}& 35.9 & 35.5 & 44.3 & 80.8& 83.6  & 32.1 & 42.6 & 827.2\\
\rowcolor{eng} wav2vec2-large \cite{baevski2020wav2vec}& 35.4 & 35.7 & 43.9 & 8.0& 78.2  & 34.7 & 42.2 & 586.9\\
\rowcolor{eng} robust-wav2vec2-large \cite{hsu21_interspeech} & 35.7 & 31.1 & 42.2 & 72.1 & 62.9 & 33.7 & 46.0 & 768.6\\
\rowcolor{euro} wav2vec2-base-23 \cite{wang2021voxpopuli} & 35.1 & 32.0 & 42.2 & 71.9 & 66.3 & 30.9 & 43.0 & 798.0 \\
\rowcolor{euro} wav2vec2-large-23 \cite{wang2021voxpopuli} & 34.2  & 35.3 & 42.4 & 64.2 & 49.7& 35.2 & 43.1 & 724.9 \\
\rowcolor{multi} XLSR-53 \cite{conneau2020unsupervised} & 34.9 & 26.9 & 40.6 & 87.1 & 76.9 & 28.6 & 44.6 & 894.0 \\
\rowcolor{multi} XLSR-128 \cite{babu2021xls} & \textbf{30.6} & \textbf{22.0} & \textbf{39.3} & \textbf{87.9} & 85.6  & \textbf{22.9} & 42.4 & \textbf{996.0}\\
\rowcolor{eng} HuBERT-base \cite{hsu2021hubert} & 35.3 & 31.4 & 42.7 & 86.1 & 86.0 & 30.9 & \textbf{41.8} & 884.9 \\
\rowcolor{eng} HuBERT-large \cite{hsu2021hubert} & 32.2 & 37.7 & 43.5 & 64.1 & 77.7 & 35.1 & 42.2 & 783.6 \\
\rowcolor{cmn} HuBERT-base-cmn \cite{mandarin_hubert}  & 35.6 & 43.2 & 46.6 & 85.3 & \textbf{86.1} & 31.8 & 42.1 & 810.2 \\
\rowcolor{cmn} HuBERT-large-cmn \cite{mandarin_hubert} & 33.7 & 39.6 & 45.1 & 57.3 & 75.6 & 37.1 & 44.4 & 713.2 \\
\rowcolor{euro} mHuBERT-base \cite{lee2022textless} & 33.0 & 33.4 & 43.6 & 72.5 & 70.9 & 29.7 & 43.1 & 812.7 \\
\bottomrule
\end{tabular}
}

    \label{tab: 1h}
        \vspace{-15pt}
\end{table*}

\subsection{Experimental Results}
\label{ssec: exp result}

The experimental results are shown in Table~\ref{tab: 10min} for 10-minute set and Table~\ref{tab: 1h} for 1-hour set. 

\noindent \textbf{Monolingual ASR}: In the monolingual ASR task, all speech SSL models outperform the FBANK baseline. XLSR-128 achieves the best performance in the 1-hour set, while HuBERT-large obtains the best performance in the 10-minute set. Several findings are noteworthy: (1) HuBERT-based models outperform wav2vec2-based models when the training data and model size are similar.
(2) Large models usually obtain better results than their base versions.
(3) While the XLSR series of models deliver impressive performances in the 1-hour set, we have observed their instability in the 10-minute set, particularly on Asian languages such as \texttt{cmn}.

\noindent \textbf{Multilingual ASR}: In the multilingual ASR task, all models trained using self-supervised learning (SSL) techniques have shown superior performance compared to the baseline model using FBANK features. Among the SSL models, XLSR-128 achieves the best results across all conditions. Our experiments also reveal some interesting findings:
(1) Models trained with more languages generally outperform those trained on monolingual datasets, although this may not always be the case. For example, mHuBERT-base performs worse than HuBERT-based models trained on English only.
(2) Large models trained on monolingual data do not necessarily have better representations for multilingual scenarios. For instance, HuBERT-large performs worse than HuBERT-base, and wav2vec2-large is less effective than wav2vec2-base. One possible explanation for the lack of performance improvement with larger models is their limited ability to generalize, despite having similar training losses as base models.
(3) The robust-wav2vec2-large model achieves decent scores on multilingual ASR, suggesting that our benchmark corpus may need to consider different acoustic environments, as it includes multiple source datasets.

\noindent \textbf{LID}: In the LID task, we notice similarities with multilingual ASR, but there are also notable differences. (1) XLSR-128 has been the dominant model for both 10-minute and 1-hour datasets. (2) While most SSL models have improvements over FBANK, some do not, particularly those based on wav2vec2 (e.g., wav2vec2-large-23 for the 10-minute set and wav2vec2-large for the 1-hour set). (3) Larger models with more parameters and pre-trained data do not necessarily lead to better performance compared to base models.

\noindent \textbf{Joint Multilingual ASR + LID}: In the joint multilingual ASR+LID task, the results generally align with the other two tasks in the multilingual track. (1) SSL models outperform FBANK on ASR, but some models perform worse on LID. (2) Base models exhibit better generalization ability and often perform better on test sets. (3) There is no single best model that dominates the task, particularly in few-shot cases and LID tasks.

\noindent \textbf{Overall}: In terms of overall performance as measured by SUPERB$_s$ in Sec.~\ref{ssec: framework}, XLSR-128 is the best model for both the 10-minute and 1-hour sets. Major findings include: (1) multilingual training with a broad coverage of languages, as seen in XLSR models that include more than 50 languages, has proven to be useful. However, multilingual training that is limited to a few selective languages may not be as beneficial in larger language groups (e.g., wav2vec2-large-23 and mHUBERT models do not always perform better than their models trained in a single language). (2) The base models tend to generalize better to multilingual cases than their corresponding large versions, such as wav2vec2-base versus wav2vec2-large and HuBERT-base versus HuBERT-large.

% \begin{itemize}
%     \item SSL can outperform FBANK on the benchmark
%     \item Model trained with more languages tends to be better than the one trained in monolingual (not exact, mhubert on 3 langs worse than hubert-eng in some cases)
%     \item Large model on monolingual may not necessarily get better results in multilingual cases (though still better in single language cases)
%     \item robustness may not be the primary concern in multilingual training (e.g., robust wav2vec2 is not helping)
% \end{itemize}

\begin{figure}[tbp]
    % \vspace{-10pt}
    \centering
    \includegraphics[width=0.9\linewidth]{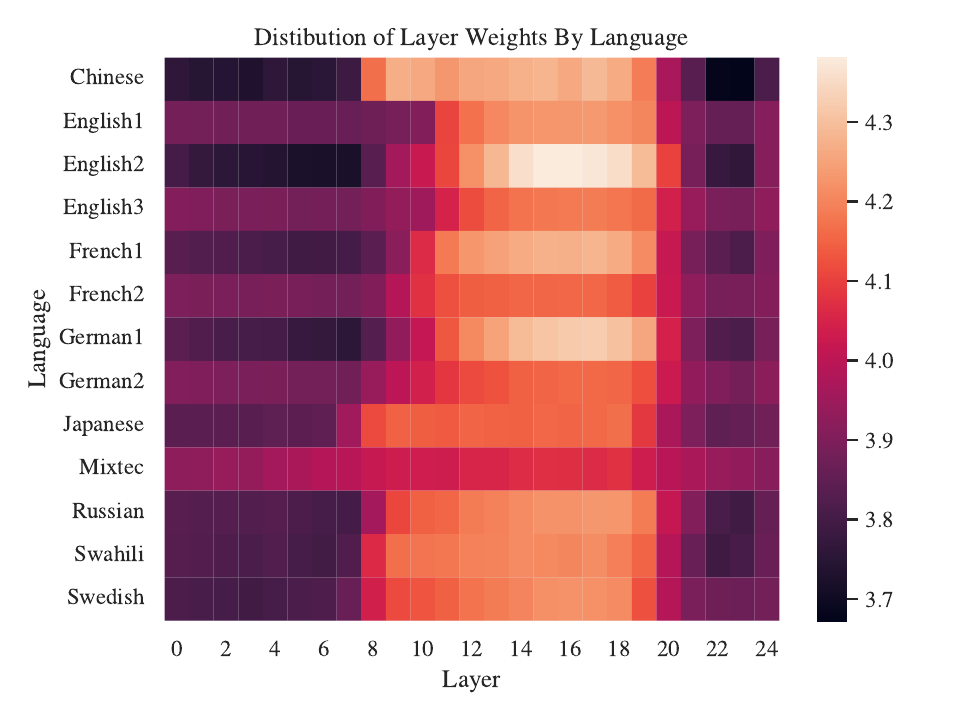}
    \vspace{-15pt}
    \caption{The layerwise weight analysis of XLSR-128 model in the monolingual track.}
    \label{fig:diagram_weight}
    \vspace{-20pt}
\end{figure}

\subsection{Layerwise analysis}

Our benchmark offers tools to guide users in the use of SSL representations according to their needs, including an analysis of the learned weights for layer importance.
The results for the XLSR-128 model in monolingual ASR tasks (shown in Fig~\ref{fig:diagram_weight}) confirm the conclusions reached by \cite{pasad2022comparative} and \cite{berrebbi2022overthinking}: the most relevant layers for ASR are not the last few layers. 
We also observed that English3, French2, and German2 have very similar behavior. These tasks use VoxPopuli data for training, which is the only dataset with lecture speech in our collection. 
Additionally, Mixtec is the only conversational speech data among our sets, and we can see a distinct behavior in Fig~\ref{fig:diagram_weight}. Therefore, the relevance of SSL model layers may be related to the speech domain (in addition to the speech task) rather than the language.

\section{Conclusion}
This paper introduces ML-SUPERB, a benchmark that extends SUPERB to multilingual tasks. 
We present the design of the open-source framework and discuss experimental results for some example models. 
More detailed policies can be found at \url{https://multilingual.superbbenchmark.org/}. 
We invite the community to participate in this challenge.

% \section{Acknowledgements}

% \ifinterspeechfinal
%      The INTERSPEECH 2023 organisers
% \else
%      The authors
% \fi
% would like to thank ISCA and the organising committees of past INTERSPEECH conferences for their help and for kindly providing the previous version of this template.

% \bibliographystyle{IEEEtran}
% \bibliography{mybib}

\section{References}
{
\printbibliography

@article{zhao2022improving,
  title={Improving Automatic Speech Recognition Performance for Low-Resource Languages With Self-Supervised Models},
  author={Zhao, Jing and Zhang, Wei-Qiang},
  journal={JSTSP},
  volume={16},
  number={6},
  pages={1227--1241},
  year={2022},
  publisher={IEEE}
}

@inproceedings{berrebbi22_interspeech,
  author={Dan Berrebbi and Jiatong Shi and Brian Yan and others},
  title={{Combining Spectral and Self-Supervised Features for Low Resource Speech Recognition and Translation}},
  year=2022,
  booktitle={Proc. Interspeech 2022},
  pages={3533--3537},
  doi={10.21437/Interspeech.2022-10796}
}

@article{wu2020self,
  title={Self-Supervised Representations Improve End-to-End Speech Translation},
  author={Wu, Anne and Wang, Changhan and Pino, Juan and Gu, Jiatao},
  journal={Proc. Interspeech 2020},
  pages={1491--1495},
  year={2020}
}

@article{conneau2020unsupervised,
  title={Unsupervised cross-lingual representation learning for speech recognition},
  author={Conneau, Alexis and Baevski, Alexei and Collobert, Ronan and Mohamed, Abdelrahman and Auli, Michael},
  journal={arXiv preprint arXiv:2006.13979},
  year={2020}
}

@article{babu2021xls,
  title={{XLS-R}: Self-supervised cross-lingual speech representation learning at scale},
  author={Babu, Arun and Wang, Changhan and Tjandra, Andros and Lakhotia, Kushal and Xu, Qiantong and Goyal, Naman and Singh, Kritika and von Platen, Patrick and Saraf, Yatharth and Pino, Juan and others},
  journal={arXiv preprint arXiv:2111.09296},
  year={2021}
}

@article{duquenne2022speechmatrix,
  title={SpeechMatrix: A Large-Scale Mined Corpus of Multilingual Speech-to-Speech Translations},
  author={Duquenne, Paul-Ambroise and Gong, Hongyu and Dong, Ning and Du, Jingfei and Lee, Ann and Goswani, Vedanuj and Wang, Changhan and Pino, Juan and others},
  journal={arXiv preprint arXiv:2211.04508},
  year={2022}
}

@article{mohamed2022self,
  title={Self-supervised speech representation learning: A review},
  author={Mohamed, Abdelrahman and Lee, Hung-yi and Borgholt, Lasse and Havtorn, Jakob D and Edin, Joakim and Igel, Christian and Kirchhoff, Katrin and Li, Shang-Wen and Livescu, Karen and others},
  journal={JSTSP},
  year={2022},
  publisher={IEEE}
}

@article{baevski2020wav2vec,
  title={wav2vec 2.0: A framework for self-supervised learning of speech representations},
  author={Baevski, Alexei and Zhou, Yuhao and Mohamed, Abdelrahman and Auli, Michael},
  journal={Proc. NeurIPS},
  volume={33},
  pages={12449--12460},
  year={2020}
}

@article{hsu2021hubert,
  title={Hu{BERT}: Self-supervised speech representation learning by masked prediction of hidden units},
  author={Hsu, Wei-Ning and Bolte, Benjamin and Tsai, Yao-Hung Hubert and Lakhotia, Kushal and Salakhutdinov, Ruslan and Mohamed, Abdelrahman},
  journal={TASLP},
  volume={29},
  pages={3451--3460},
  year={2021},
  publisher={IEEE}
}

@inproceedings{yang21c_interspeech,
  author={Shu-wen Yang and Po-Han Chi and Yung-Sung Chuang and Cheng-I Jeff Lai and Kushal Lakhotia and Yist Y. Lin and Andy T. Liu and Jiatong Shi and Xuankai Chang and Guan-Ting Lin and Tzu-Hsien Huang and Wei-Cheng Tseng and Ko-tik Lee and Da-Rong Liu and Zili Huang and Shuyan Dong and Shang-Wen Li and Shinji Watanabe and Abdelrahman Mohamed and Hung-yi Lee},
  title={{SUPERB: Speech Processing Universal PERformance Benchmark}},
  year=2021,
  booktitle={Proc. Interspeech 2021},
  pages={1194--1198},
  doi={10.21437/Interspeech.2021-1775}
}

@inproceedings{tsai2022superb,
  title={{SUPERB-SG}: Enhanced Speech processing Universal PERformance Benchmark for Semantic and Generative Capabilities},
  author={Tsai, Hsiang-Sheng and Chang, Heng-Jui and Huang, Wen-Chin and Huang, Zili and Lakhotia, Kushal and Yang, Shu-wen and Dong, Shuyan and Liu, Andy and Lai, Cheng-I and Shi, Jiatong and others},
  booktitle={Proceedings of the 60th Annual Meeting of the Association for Computational Linguistics (Volume 1: Long Papers)},
  pages={8479--8492},
  year={2022}
}

@inproceedings{evain21_interspeech,
  author={Sol{\'e}ne Evain and Ha Nguyen and Hang Le and Marcely Zanon Boito and Salima Mdhaffar and Sina Alisamir and Ziyi Tong and Natalia Tomashenko and Marco Dinarelli and Titouan Parcollet and Alexandre Allauzen and others},
  title={{ LeBenchmark: A Reproducible Framework for Assessing Self-Supervised Representation Learning from Speech}},
  year=2021,
  booktitle={Proc. Interspeech 2021},
  pages={1439--1443},
  doi={10.21437/Interspeech.2021-556}
}

@inproceedings{conneau22_interspeech,
  author={Alexis Conneau and Ankur Bapna and Yu Zhang and Min Ma and Patrick {von Platen} and Anton Lozhkov and Colin Cherry and Ye Jia and Clara Rivera and Mihir Kale and Daan {van Esch} and Vera Axelrod and Simran Khanuja and Jonathan Clark and Orhan Firat and Michael Auli and Sebastian Ruder and Jason Riesa and Melvin Johnson},
  title={{XTREME-S: Evaluating Cross-lingual Speech Representations}},
  year=2022,
  booktitle={Proc. Interspeech 2022},
  pages={3248--3252},
  doi={10.21437/Interspeech.2022-10007}
}

@article{pratap2020mls,
  title={{MLS}: A Large-Scale Multilingual Dataset for Speech Research},
  author={Pratap, Vineel and Xu, Qiantong and Sriram, Anuroop and Synnaeve, Gabriel and Collobert, Ronan},
  journal={Proc. Interspeech 2020},
  pages={2757--2761},
  year={2020}
}

@inproceedings{ardila2020common,
  title={Common Voice: A Massively-Multilingual Speech Corpus},
  author={Ardila, Rosana and Branson, Megan and Davis, Kelly and Kohler, Michael and Meyer, Josh and Henretty, Michael and Morais, Reuben and Saunders, Lindsay and Tyers, Francis and Weber, Gregor},
  booktitle={Proc. LREC},
  pages={4218--4222},
  year={2020}
}

@article{maclean2018voxforge,
  title={Voxforge},
  author={MacLean, Ken},
  journal={Ken MacLean.[Online]. Available: http://www. voxforge. org/home.[Accessed by 2022]},
  year={2018}
}

@inproceedings{wang2021voxpopuli,
  title={Vox{P}opuli: A Large-Scale Multilingual Speech Corpus for Representation Learning, Semi-Supervised Learning and Interpretation},
  author={Wang, Changhan and Riviere, Morgane and Lee, Ann and Wu, Anne and Talnikar, Chaitanya and Haziza, Daniel and Williamson, Mary and Pino, Juan and Dupoux, Emmanuel},
  booktitle={Proceedings of the 59th Annual Meeting of the Association for Computational Linguistics and the 11th International Joint Conference on Natural Language Processing (Volume 1: Long Papers)},
  pages={993--1003},
  year={2021}
}

@inproceedings{kjartansson-etal-tts-sltu2018,
    title = {A Step-by-Step Process for Building TTS Voices Using Open Source Data and Framework for Bangla, Javanese, Khmer, Nepali, Sinhala, and Sundanese},
    author = {Keshan Sodimana and Knot Pipatsrisawat and Linne Ha and Martin Jansche and Oddur Kjartansson and Pasindu De Silva and Supheakmungkol Sarin},
    booktitle = {Proc. SLTU},
    year  = {2018},
    pages = {66--70},
}

@inproceedings{kjartansson-etal-2020-open,
    title = {Open-Source High Quality Speech Datasets for Basque, Catalan and Galician},
    author = {Kjartansson, Oddur and Gutkin, Alexander and Butryna, Alena and Demirsahin, Isin and Rivera, Clara},
    booktitle = {Proc. SLTU},
    year = {2020},
    pages = {21--27},
}

@inproceedings{he-etal-2020-open,
    title = {Open-source Multi-speaker Speech Corpora for Building Gujarati, Kannada, Malayalam, Marathi, Tamil and Telugu Speech Synthesis Systems},
    author = {He, Fei and Chu, Shan-Hui Cathy and Kjartansson, Oddur and Rivera, Clara and Katanova, Anna and Gutkin, Alexander and Demirsahin, Isin and Johny, Cibu and Jansche, Martin and Sarin, Supheakmungkol and Pipatsrisawat, Knot},
    booktitle = {Proc. LREC},
    year = {2020},
    pages = {6494--6503},
}

@incollection{rehm2012language,
  title={Language Technology Support for Norwegian},
  author={Rehm, Georg and Uszkoreit, Hans},
  booktitle={The Norwegian Language in the Digital Age: Bokmalsversjon},
  pages={52--70},
  year={2012},
  publisher={Springer}
}

@inproceedings{conneau2023fleurs,
  title={Fleurs: Few-shot learning evaluation of universal representations of speech},
  author={Conneau, Alexis and Ma, Min and Khanuja, Simran and Zhang, Yu and Axelrod, Vera and Dalmia, Siddharth and Riesa, Jason and Rivera, Clara and Bapna, Ankur},
  booktitle={2022 IEEE Spoken Language Technology Workshop (SLT)},
  pages={798--805},
  year={2023},
  organization={IEEE}
}

@inproceedings{barnard2014nchlt,
  title={The NCHLT speech corpus of the South African languages},
  author={Barnard, Etienne and Davel, Marelie H and van Heerden, Charl and De Wet, Febe and Badenhorst, Jaco},
  year={2014},
  organization={Proc. SLTU}
}

@article{baumann2019spoken,
  title={The Spoken Wikipedia Corpus collection: Harvesting, alignment and an application to hyperlistening},
  author={Baumann, Timo and K{\"o}hn, Arne and Hennig, Felix},
  journal={LREC},
  volume={53},
  pages={303--329},
  year={2019},
  publisher={Springer}
}

@inproceedings{shi2021leveraging,
  title={Leveraging End-to-End ASR for Endangered Language Documentation: An Empirical Study on Yol{\'o}xochitl Mixtec},
  author={Shi, Jiatong and others},
  booktitle={Proceedings of the 16th Conference of the European Chapter of the Association for Computational Linguistics: Main Volume},
  pages={1134--1145},
  year={2021}
}

@inproceedings{shi2021highland,
  title={Highland Puebla Nahuatl speech translation corpus for endangered language documentation},
  author={Shi, Jiatong and Amith, Jonathan D and Chang, Xuankai and Dalmia, Siddharth and Yan, Brian and Watanabe, Shinji},
  booktitle={Proc. AmericaNLP},
  pages={53--63},
  year={2021}
}

@article{mailab,
  title={M-AILab Speech dataset},
  author={Imdat Solak},
  journal={Imdat Solak.[Online]. Available: https://www.caito.de/2019/01/03/the-m-ailabs-speech-dataset/.[Accessed by 2022]},
  year={2018}
}

@inproceedings{braude2019all,
  title={All Together Now: The Living Audio Dataset.},
  author={Braude, David A and Aylett, Matthew P and Laoide-Kemp, Caoimh{\'\i}n and Ashby, Simone and Scott, Kristen M and Raghallaigh, Brian O and Braudo, Anna and Brouwer, Alex and Stan, Adriana},
  booktitle={INTERSPEECH},
  pages={1521--1525},
  year={2019}
}

@article{de2014smartphone,
  title={A smartphone-based ASR data collection tool for under-resourced languages},
  author={De Vries, Nic J and Davel, Marelie H and Badenhorst, Jaco and Basson, Willem D and De Wet, Febe and Barnard, Etienne and De Waal, Alta},
  journal={Speech communication},
  volume={56},
  pages={119--131},
  year={2014},
  publisher={Elsevier}
}

@inproceedings{watanabe2018espnet,
  author={Shinji Watanabe and Takaaki Hori and Shigeki Karita and Tomoki Hayashi and Jiro Nishitoba and Yuya Unno and Nelson {Enrique Yalta Soplin} and Jahn Heymann and Matthew Wiesner and Nanxin Chen and Adithya Renduchintala and Tsubasa Ochiai},
  title={{ESPnet}: End-to-End Speech Processing Toolkit},
  year={2018},
  booktitle={Proceedings of Interspeech},
  pages={2207--2211},
  doi={10.21437/Interspeech.2018-1456},
  url={http://dx.doi.org/10.21437/Interspeech.2018-1456}
}

@inproceedings{wolf-etal-2020-transformers,
    title = "Transformers: State-of-the-Art Natural Language Processing",
    author = "Wolf, Thomas  and
      Debut, Lysandre  and
      Sanh, Victor  and
      Chaumond, Julien  and
      Delangue, Clement  and
      Moi, Anthony  and
      Cistac, Pierric  and
      Rault, Tim  and
      Louf, Remi  and
      Funtowicz, Morgan  and
      Davison, Joe  and
      Shleifer, Sam  and
      von Platen, Patrick  and
      Ma, Clara  and
      Jernite, Yacine  and
      Plu, Julien  and
      Xu, Canwen  and
      Le Scao, Teven  and
      Gugger, Sylvain  and
      Drame, Mariama  and
      Lhoest, Quentin  and
      Rush, Alexander",
    booktitle = "Proceedings of the 2020 Conference on Empirical Methods in Natural Language Processing: System Demonstrations",
    month = oct,
    year = "2020",
    address = "Online",
    publisher = "Association for Computational Linguistics",
    url = "https://aclanthology.org/2020.emnlp-demos.6",
    doi = "10.18653/v1/2020.emnlp-demos.6",
    pages = "38--45",
}

@inproceedings{feng2023superb,
  title={{SUPERB@ SLT} 2022: Challenge on Generalization and Efficiency of Self-Supervised Speech Representation Learning},
  author={Feng, Tzu-hsun and Dong, Annie and Yeh, Ching-Feng and Yang, Shu-wen and Lin, Tzu-Quan and Shi, Jiatong and Chang, Kai-Wei and Huang, Zili and Wu, Haibin and Chang, Xuankai and others},
  booktitle={2022 IEEE Spoken Language Technology Workshop (SLT)},
  pages={1096--1103},
  year={2023},
  organization={IEEE}
}

@inproceedings{chang2021exploration,
  title={An exploration of self-supervised pretrained representations for end-to-end speech recognition},
  author={Chang, Xuankai and Maekaku, Takashi and Guo, Pengcheng and Shi, Jing and Lu, Yen-Ju and Subramanian, Aswin Shanmugam and Wang, Tianzi and Yang, Shu-wen and Tsao, Yu and Lee, Hung-yi and others},
  booktitle={2021 IEEE Automatic Speech Recognition and Understanding Workshop (ASRU)},
  pages={228--235},
  year={2021},
  organization={IEEE}
}

@article{chen2022wavlm,
  title={Wav{LM}: Large-scale self-supervised pre-training for full stack speech processing},
  author={Chen, Sanyuan and Wang, Chengyi and Chen, Zhengyang and Wu, Yu and Liu, Shujie and Chen, Zhuo and Li, Jinyu and Kanda, Naoyuki and Yoshioka, Takuya and Xiao, Xiong and others},
  journal={JSTSP},
  volume={16},
  number={6},
  pages={1505--1518},
  year={2022},
  publisher={IEEE}
}

@inproceedings{lee2022textless,
  title={Textless Speech-to-Speech Translation on Real Data},
  author={Lee, Ann and Gong, Hongyu and Duquenne, Paul-Ambroise and Schwenk, Holger and Chen, Peng-Jen and Wang, Changhan and Popuri, Sravya and Adi, Yossi and Pino, Juan and Gu, Jiatao and others},
  booktitle={Proceedings of the 2022 Conference of the North American Chapter of the Association for Computational Linguistics: Human Language Technologies},
  pages={860--872},
  year={2022}
}

@online{mandarin_hubert,
  author = {Liu, Shixing and Guo, Pengcheng},
  title = {Chinese speech pretraining},
  year = 2023,
  url = {https://github.com/TencentGameMate/chinese_speech_pretrain},
  urldate = {2022-06-30}
}

@inproceedings{hsu21_interspeech,
  author={Wei-Ning Hsu and Anuroop Sriram and Alexei Baevski and Tatiana Likhomanenko and Qiantong Xu and Vineel Pratap and Jacob Kahn and Ann Lee and Ronan Collobert and Gabriel Synnaeve and Michael Auli},
  title={{Robust wav2vec 2.0: Analyzing Domain Shift in Self-Supervised Pre-Training}},
  year=2021,
  booktitle={Proc. Interspeech 2021},
  pages={721--725},
  doi={10.21437/Interspeech.2021-236}
}

@inproceedings{hou20_interspeech,
  author={Wenxin Hou and Yue Dong and Bairong Zhuang and Longfei Yang and Jiatong Shi and Takahiro Shinozaki},
  title={{Large-Scale End-to-End Multilingual Speech Recognition and Language Identification with Multi-Task Learning}},
  year=2020,
  booktitle={Proc. Interspeech 2020},
  pages={1037--1041},
  doi={10.21437/Interspeech.2020-2164}
}

@inproceedings{watanabe2017language,
  title={Language independent end-to-end architecture for joint language identification and speech recognition},
  author={Watanabe, Shinji and Hori, Takaaki and Hershey, John R},
  booktitle={2017 IEEE Automatic Speech Recognition and Understanding Workshop (ASRU)},
  pages={265--271},
  year={2017},
  organization={IEEE}
}

@inproceedings{zhang22da_interspeech,
  author={Chao Zhang and Bo Li and Tara Sainath and Trevor Strohman and Sepand Mavandadi and Shuo-Yiin Chang and Parisa Haghani},
  title={{Streaming End-to-End Multilingual Speech Recognition with Joint Language Identification}},
  year=2022,
  booktitle={Proc. Interspeech 2022},
  pages={3223--3227},
  doi={10.21437/Interspeech.2022-11249}
}

@article{chen2023improving,
  title={Improving Massively Multilingual {ASR} With Auxiliary {CTC} Objectives},
  author={Chen, William and Yan, Brian and Shi, Jiatong and Peng, Yifan and Maiti, Soumi and Watanabe, Shinji},
  journal={Proc. ICASSP 2023},
  year={2023}
}

@article{pasad2022comparative,
  title={Comparative layer-wise analysis of self-supervised speech models},
  author={Pasad, Ankita and Shi, Bowen and Livescu, Karen},
  journal={Proc. ICASSP 2023},
  year={2022}
}

@article{berrebbi2022overthinking,
  title={Avoid Overthinking in Self-Supervised Models for Speech Recognition},
  author={Dan Berrebbi and Brian Yan and Shinji Watanabe},
  journal={Proc. ICASSP 2023},
  year={2022}
}

@inproceedings{li22aa_interspeech,
  author={Xinjian Li and Florian Metze and David R. Mortensen and Alan W Black and Shinji Watanabe},
  title={{ASR2K: Speech Recognition for Around 2000 Languages without Audio}},
  year=2022,
  booktitle={Proc. Interspeech 2022},
  pages={4885--4889},
  doi={10.21437/Interspeech.2022-10712}
}

@article{javed2022indicsuperb,
  title={IndicSUPERB: A Speech Processing Universal Performance Benchmark for Indian languages},
  author={Javed, Tahir and Bhogale, Kaushal Santosh and Raman, Abhigyan and Kunchukuttan, Anoop and Kumar, Pratyush and Khapra, Mitesh M},
  journal={arXiv preprint arXiv:2208.11761},
  year={2022}
}

@misc{amith_yoloxochitl_mixtec,
  author = {Amith, Jonathan D. and Castillo Castillo, Rey},
  title = {Audio corpus of Yoloxóchitl Mixtec with accompanying time-coded transcriptons in ELAN},
  note = {n.d.},
url={https://www.openslr.org/89/}
}

@misc{amith_audio_corpus_sierra,
  author = {Amith, Jonathan D. and Domínguez Alcántara, Amelia and Salazar Osollo, Hermelindo and Salgado Castañeda, Ceferino and Gorostiza Salazar, Eleuterio},
  title = {Audio corpus of Sierra Nororiental and Sierra Norte de Puebla Nahuat(l) with accompanying time-code transcriptions in ELAN},
  note = {n.d.},
url={https://www.openslr.org/92/}
}

@misc{amith_totonac,
  author = {Amith, Jonathan D. and López Francisco, Osbel},
  title = {Audio corpus of Totonac recordings from northern Puebla and adjacent areas of Veracruz},
  note = {n.d.},
  url = {https://www.openslr.org/107/}
}
}

\end{document}